\documentstyle[twocolumn,aps,epsfig]{revtex}
\tighten
\begin{document}
\draft

\title{Te covered Si(001): a variable surface reconstruction}

\author{Prasenjit Sen$^{(1)}$, S. Ciraci$^{(1,2)}$, 
Inder P. Batra$^{(1)}$, and C. H. Grein$^{(1)}$}

\address{$^{(1)}$ Department of Physics, University of Illinois at Chicago,
Chicago, IL 60607-7059}
\address{$^{(2)}$ Department of Physics, Bilkent University, Bilkent,Ankara, 
06533 Turkey}

\address{
\begin{minipage}[t]{6.0in}
\begin{abstract}
At a given temperature, clean and adatom covered silicon surfaces usually
exhibit well-defined reconstruction patterns. Our finite temperature
ab-initio molecular dynamics calculations show that the tellurium covered
Si(001) surface is an exception. Soft longitudinal modes of surface 
phonons due to the strongly anharmonic potential of the bridged 
tellurium atoms prevent the reconstruction structure from attaining any 
permanent, two dimensional periodic geometry. This explains why 
experiments attempting to find a definite model for the reconstruction 
have reached conflicting conclusions.
\typeout{polish abstract}
\end{abstract}
\end{minipage}}
\date{\today}
\maketitle


Relaxation and reconstruction of clean and adatom covered 
surfaces is an active field of study. Tremendous efforts have been
devoted to observing and understanding how the symmetry and atomic 
configurations of surfaces change, and how these changes affect the
chemical and physical properties of surfaces. Particular atomic
structures with well-defined reconstruction geometry are verified
and sometimes predicted,
by performing static total energy calculations at T=0 K. 
Atomic configurations corresponding to the global or local minima on a
Born-Oppenheimer surface are then attributed to stable surface structures.

In an effort to promote technology by growing crystals with the minimum
possible defects, the clean and adatom covered surfaces of silicon 
have been thoroughly investigated. Recent studies on the growth of 
silicon surfaces have shown that atoms like As, Sb, Te are
good surfactants preventing island formation and hence aiding the
layer by layer growth.\cite{copel0,tromp0,higuch,grandj,benne1,fong00}  
In addition, the goal of
combining large infra-red detector arrays with relatively cheap and
well developed Si integrated circuit technology has led to the growth of
HgCdTe on Si(001) surface. Because of a large ($\sim 19\%$) lattice mismatch,
thick buffers layers of CdTe must be grown before growing active 
layers of HgCdTe.\cite{benne2} The atomic configuration of adsorbed
Te, which forms the first layer grown on the bare Si(001) surface, 
is crucial for the fabrication of high performance devices, since
it determines how growth nucleates and hence the quality of the 
epilayer.\cite{almeida}

The adsorption of Te on Si(001) surfaces for different 
coverages and the resulting atomic geometries have been studied 
by different surface techniques. While the energetically favorable 
adsorption sites are well understood at low Te 
coverage,\cite{benne2,quate0} the models of surface reconstructions
proposed for near monolayer coverage ($\Theta \sim $1) have been at 
variance.\cite{nardo,tamiya,wiame0} Most of the LEED experiments observe
a (1$\times$1) structure up to temperatures high enough to desorb 
Te from the surface.\cite{nardo} Other experiments report 
different structures. For example, Tamiya {\it et al.}\cite{tamiya} 
found the transition from the low temperature (1$\times$1) 
structure to a (2$\times$1) structure at T=873 K
Wiame {\it et al.}\cite{wiame0} observed a (2$\times$1) symmetry in their
STM images even at room temperature and proposed an atomistic model 
involving Te-Te dimers. The differing of the reconstruction 
geometries from one experiment to another is a puzzling and 
an uncommon situation.   

In this work, we explain this puzzling situation using
the results of finite temperature ab-initio molecular dynamics (MD)
calculations. We investigate the energetics of Te adsorption starting
from very low coverage ($\Theta=$0.0625) up to a monolayer coverage 
($\Theta=$1). We first determine the binding energies of a single Te 
atom adsorbed at the special sites of the unit cell for $\Theta \ll 1$. 
We describe how the original Si--Si dimer bonds of the
Si(001)-(2$\times$1) surface are broken, and how eventually the surface is 
covered by Te atoms. We also examine the possibility 
of two adjacent adsorbed Te atoms forming a dimer bond to give a  
(2$\times$1) reconstruction. Other possible higher 
order reconstruction geometries are searched by a finite
temperature, ab initio molecular dynamics method. We find that 
uncorrelated lateral excursions of bridged Te atoms in flat potential
wells hinder the observation of any definitive surface reconstruction 
pattern at finite temperatures.

Calculations were
carried out within the density functional approach using the Vienna
Ab-initio Simulation Package (VASP)\cite{vasp00}. The wave 
functions are expressed by plane waves with the cutoff energy 
$|{\bf k}+{\bf G}|^2 \leq 250$ eV. The Brillouin zone (BZ) integration
is performed by using the Monkhorst-Pack scheme\cite{monkho}
with (2$\times$2$\times$1),  
(2$\times$8$\times$1) and (4$\times$8$\times$1) special points for 
(4$\times$4), (4$\times$1) and (2$\times$1) cells, respectively.
The convergence with respect to the energy cutoff and number of 
{\bf k}-points was tested. Ionic potentials are represented by 
ultra-soft Vanderbilt type
pseudopotentials\cite{vander} and results are obtained within generalized 
gradient approximation\cite{perdew} for a fully relaxed atomic structure.
The preconditioned conjugate gradient method is used for wave function 
optimization and the conjugate gradient method for ionic relaxation at 
$T=$0 K. At finite temperatures, the
Nos\'{e}-Hoover thermostat\cite{nose00} is employed for constant temperature
dynamics of ionic motions in the self-consistent field of 
electrons.\cite{vasp00}  The time step in MD calculations, $\Delta t$ is 
chosen such that typical phonon time period is divided into a few tens of 
time steps. We picked $\Delta t$ to be 2 $fs$ to ensure that the ionic 
trajectories are smooth. 

The Si(001) surface is represented by a repeating slab geometry. Each 
slab contains 5 Si(001) atomic planes and hydrogen atoms passivating the 
Si atoms at the bottom of the slab. Consecutive slabs are separated 
by a vacuum space of 9 \AA. For calculations at $T=$0 K, 
Si atoms in the top four atomic layers are allowed to relax, while the 
bottom Si atoms and passivating hydrogens are fixed
to simulate bulk-like termination. 

In finite temperature
calculations, all atoms, including Si and H atoms in the bottom layer, 
are allowed to move to avoid a large temperature gradient. Lattice 
parameters are expanded according to the temperature under study 
using the experimental thermal expansion coefficient in order to prevent 
the lattice from experiencing internal thermal strain.
We reproduced the energetics and geometry of the $c(4\times 2)$, 
$p(2\times 2)$ and $p(2\times 1)$ reconstructions of a clean Si(001) 
surface using the above parameters.\cite{kelly0}

The binding energy of a single Te adsorbed on the special 
(on-top {\bf T}, cave {\bf C}, hollow {\bf H}, and bridge {\bf B}) 
sites on the clean Si(001) surface are calculated using a supercell 
consisting of eight (2$\times$1) cells. The large size of the supercell 
ensures that the interaction between the adsorbed Te atoms is negligible
so that results can represent low Te coverage. In Fig. 1a, only
one (2$\times$1) cell of the supercell is shown. The binding energies
are found to be {\bf T}: 4.5 eV, {\bf C}: 3.5 eV, {\bf H}: 3.4 eV, 
and {\bf B}: 3.2 eV.
These binding energies were calculated for fully relaxed structures at 
T=0 K. Apparently, the most energetic site at low coverage is the on-top 
site, where a Te atom above the dimer bond of the clean 
Si(001)-(2$\times$1) surface is bonded to two Si atoms of the same dimer 
bond. This is consistent with our intuitive chemical notion that
Te(5$p^4$) tries to fill its outermost $p$-shell by coordinating
with two surface Si atoms.
Our result is also in agreement with STM images.\cite{quate0}
By considering only two special sites, Takeuchi\cite{takeuc} found the 
on-top site to be energetically more favorable than the bridge 
sites by 0.8 eV. We examined
the stability of the Te atom adsorbed at the on-top site for higher
coverages. For $\Theta=$0.5, Te atoms adsorbed 2.25~ \AA~ above each surface 
dimer bond           
were found stable, except that the underlying Si-Si dimer bond 
is elongated marginally and the dimer asymmetry is removed. The Si-Te bond
length is 2.53 \AA~ which is close to the sum of the Si and Te 
covalent radii and in excellent agreement with 
experiment.\cite{burge}

A monolayer coverage of Te ({\it i.e.} $\Theta=$1) is the most 
critical insofar as the controversy regarding the surface 
reconstruction is concerned. We attempt to resolve the controversy by
addressing the following issues which are not settled yet. These are:
i) How does the atomic configuration of the surface change with increasing 
$\Theta >$0.5 ? ii) Can two adjacent Te atoms on the surface dimerize at 
$\Theta \sim$ 1? iii) What is the geometry of the surface reconstruction 
and how does the surface structure vary with 
temperature at $\Theta \sim$ 1? To address the first question,
we begin with an initial configuration where one Te is adsorbed at the
{\bf T}-site and the second one at the {\bf B} site on the
Si(001)-(2$\times$1) surface, and let this structure 
relax at T=0 K. The occupation of the {\bf B} site at high Te coverage is
consistent with experiments.\cite{benne2,coads} 
In reaching the stable structure, the Te atoms form directional bonds
with surface Si atoms while Si-Si dimer bonds elongate and eventually  
break. It appears that each Si-Si dimer bond is broken to form four 
new Si-Te bonds. In the final stable structure, Si atoms of the broken 
dimer bond are pushed to their bulk positions reforming the outermost, 
bulklike Si(001) atomic plane. Each adsorbed Te atom is connected 
to the substrate with two Te-Si bonds of length 2.53 \AA. At the end, 
a metallic Te(001) atomic plane forms 1.65 \AA~ above the Si substrate 
with a binding energy of 4.28 eV per Te atom relative to the clean 
Si(001)-(2$\times$1) surface and free Te atom. Figure 1b describes
the atomic positions of this ideal (1$\times$1) structure of the
Te monolayer on the Si surface. The charge density contour plots in
Fig. 1c indicate that the bond is directional. The maximum of the 
charge occurs between Si and Te, but closer to Te.\cite{ionic} 
In spite of the directional Te-Si bonds, the surface of 
Te covered Si(001) surface is metallic with a finite density of states
at the Fermi level. From force calculations we find that the Te 
atoms are robust against displacements along the [110] (or $x$-) 
direction in the plane of Si-Te-Si bonds. 

Our calculations for a free Te$_{2}$ molecule predict a binding energy of
4.41 eV and a bond length of 2.56 \AA. This suggests the possibility that 
two adjacent Te atoms on the Si(001) surface may experience energy benefit
by forming a Te-Te dimer bond by moving towards each other in the 
$y$-direction. (See Fig. 1b) Such a dimerization can 
once again lead to a (2$\times$1) reconstruction. As a matter of fact, a 
similar As-As dimerization is known to occur on As covered Si(001) 
and Ge(001) surfaces.\cite{tromp0,thorpe} To test whether Te-Te dimerization 
can occur and to answer point ($ii$), an initial structure with 
a Te-Te distance of 3 \AA~ (which is greater than the bond length 
of Te$_{2}$, but smaller than the undimerized distance in the 
(1$\times$1) structure) is relaxed at T=0 K. Upon relaxation, Te atoms 
moved away from each other so that the tilted Si-Te-Si plane became 
perpendicular to the surface and the total energy of the system is 
lowered significantly. The analysis of the charge density in a (001) (or 
$xy$) plane passing through the Te atoms suggests that the formation of 
strong Te-Si bonds excludes the bonding between two adjacent Te 
atoms (Fig. 1d). Simple valence arguments also suggests that Te being 
divalent, would tend to avoid bonding with three other atoms.

 To address the most significant question ($iii$) posed above, one 
must consider the reconstruction at $\Theta =$1 which may involve  
complex and concerted rearrangements of the
substrate and adsorbate atoms at high temperature. 
To access all possible reconstruction geometries
that cannot be easily determined by transition 
state analysis at T=0 K,  we performed a finite temperature, 
ab-initio molecular dynamics calculations at T=600 K and T=1000 K 
using (4$\times$1) supercell geometry.\cite{recons}
Figure 2 illustrates the displacements of Te atoms in a (4$\times$1)
supercell at T=600 K. The time variation of the mean squared planar 
displacements, 
$\langle u_{\parallel}^2 \rangle
=\frac{1}{4}\sum_{i=1}^{4}(u_{x,i}^2+u_{y,i}^2)$, 
($u$'s are the displacements of the atoms from their ideal lattice
positions)
shows that the system is sufficiently thermalized 
within  $\sim$ 1 $ps$. We note that the displacements along the
$x$-direction, $u_{x,i=1,4}(t)$, are small since the bridged
Si-Te-Si bonds are robust. The average of the perpendicular positions 
of Te atoms on the surface, 
$\langle z \rangle=\frac{1}{4}\sum_{i=1}^{4} z_i(t)$,
and also those of 8 hydrogen atoms at the bottom drifts 
along the $z$-direction with the same negative velocity, 
$d\langle z \rangle/dt \sim $-0.7 \AA/$ps$. In addition to this spurious 
translation of the unit cell, the displacement of each Te atom, 
$u_{z,i}(t)$, oscillates with decreasing amplitude and without any 
correlation with the other Te atoms.

The displacement along the [1\={1}0] (or $y$-) direction, 
$u_{y,i}(t)$, is large 
and can be relevant for a particular reconstruction structure. After the 
thermalization of the system, $u_{y,i}(t)$ becomes oscillatory and quasi
periodic with periods of the order of $\sim$ 1.0 $ps$. The behavior
illustrated in Fig. 2 is reminiscent of the surface longitudinal acoustic
mode due to Te rows. The amplitudes
of oscillations vary between 0.4 \AA~ to 0.7 \AA~ resulting in
lateral excursions (as large as 1.4 \AA) of Te rows along [1\={1}0] 
direction. To enhance the statistics, we performed the same
calculation at T=1000 K. The adsorbed Te atoms execute similar
motions, only with larger amplitudes, at this higher temperature.

These excursions or displacements of adjacent rows do not 
display any correlation. Moreover, they are time dependent.
The random and uncorrelated nature of the displacements prevents us from 
deducing a well-defined reconstruction pattern. Such excursions of Te rows 
along the [1\={1}0] direction would not give rise to any resolvable pattern
in the LEED and STM images. 
For example, since the period of oscillations are much 
shorter than the characteristic scan time of STM, the STM images 
taken at finite temperature would indicate disordered (1$\times$1) 
reconstruction.
 
 For adsorbed Te rows to execute large amplitude excursions with low 
frequency
at T=600 K is unusual and suggests rather soft and non-Hookian (nonlinear) 
force constants in this direction. In fact, as seen in Fig. 3, the total 
energy remains practically unchanged for a displacement of the Te rows 
of $u_{y} \sim \pm $ 0.5 \AA. For the
displacement of adjacent rows in opposite directions, $E_{T}(u_{y})$
resembles a double well potential with a broad maximum
at $u_{y}=$0 and a shallow minimum on either sides. The barrier between
these two minima is very low, almost at the accuracy limit of the
present calculations (7 meV). This suggests that  
adjacent Te rows are displaced 
by $\sim$0.25 \AA~ in opposite directions, forming a
zigzag chain of Te atoms on the [110] direction and leading to a 
(2$\times$1) surface reconstruction at T=0 K. Interestingly,
except for the disappearance of the weak double well form, 
the variation of the total energy with $u_{y}$ remains 
essentially unaltered if the adjacent
Te rows are displaced in the same direction. This implies that,
at finite temperatures, Te rows can easily traverse the weak 
barrier and execute 
random (uncorrelated) displacements. This situation is consistent with 
the results of finite temperature MD calculations summarized in Fig. 2.
Since the potential energy well is so flat, the positions of Te atoms
would be easily modified by the tip-sample interaction in STM experiments.
The total energy curve in Fig. 3, is a fit to an analytical form
$E_{T}(u_{y})=\alpha u_{y}^2 + \beta u_{y}^4 + \gamma u_{y}^6$, 
(with $\alpha=$0.3024 eV/\AA$^2$, $\beta=$0.6242 eV/\AA$^4$, 
$\gamma=-$0.2087 eV/\AA$^6$) and reflects strong anharmonicity 
(nonlinearity in force constants) 
of the potential wells wherein Te atoms move.\cite{hightc}
 
In summary, we have found that Te atoms adsorb above the Si-Si dimer
bonds at low coverage. There is no energy benefit for forming Te dimers 
at any coverage. At monolayer coverage, the potential wells for 
Te atoms are rather flat and strongly anharmonic along the [1\={1}0]
direction. There is almost no barrier for the Te rows on the 
surface to make significant excursions relative to their ideal 
positions along the [1\={1}0] direction. First principle finite 
temperature calculations indicate
that the displacements of Te rows are uncorrelated, lacking any 
definitive reconstruction pattern.

\begin{figure}
\caption{(a) The unit cell of the Si(001)-(2$\times$1) surface. The
	 possible sites for the adsorption of Te at very low $\Theta$
	 are marked by ${\bf X}$. (b) The (1$\times$1) structure of 
	 the Te covered
	 Si(001) surface. (c) Charge density contour plots of the Si-Te-Si 
	 bonds with  arrows showing the direction of increasing charge 
	 density. (d) Charge density contour plots on the (001) plane 
	 passing through the Te atoms. Large filled, large empty, small
	 empty, and smallest empty circles denote Te, first layer Si,
	 second layer Si, and third layer Si atoms, respectively. The
	 thick lines between circles indicate bonds. $x$-, $y$-, and
	 $z$-axis are parallel to the [110], [1\={1}0], and [001] 
	 directions, respectively. The lattice constant $a=$3.84 \AA.}
\end{figure}

\begin{figure}
\caption{Time variation of the displacements of the Te atoms ($u_{x}$, 
	 $u_{y}$, and $u_{z}$) from their ideal lattice positions
	 in a (4$\times$1) supercell calculated
	 at T=600K. The left panel shows the supercell. At t=0, all
	 the atoms are at their ideal lattice positions.} 
\end{figure}

\begin{figure}
\caption{Variation of total energy with the displacements of the
	 Te row, $u_{y}$, calculated at T=0 K. Each data point 
	 corresponds to a fully relaxed structure under a given
	 displacement of the Te rows along the [1\={1}0] direction. 
	 The thick line is for the adjacent rows 
	 moving in opposite directions forming a zigzag pattern.
	 $\Diamond$'s correspond to the Te rows moving in the same 
	 direction. An analytical fit, $E_{T}(u_{y})$, to the total energy 
	 values depicted by the $\Diamond$'s is shown by the broken line.
	 Energies are measured with respect to the perfect (1 $\times$
	 1) surface for $u_y = 0$.}
\end{figure}


\begin{references}
\bibitem{recons} We recognize that high order reconstructions cannot be
revealed by using a (4$\times$1) supercell. For the Te covered Si(001)
no high order reconstruction have been reported at $\Theta \sim 1$.
\end{references}
\end{document}